\documentstyle[aps,prl,preprint,tighten]{revtex}
\begin{document}
\title {NMR relaxation in half-integer
antiferromagnetic spin chains}
\author{\large Subir Sachdev}
\address{Department of Physics, P.O.
Box 208120, \\ Yale University, New Haven, CT 06520-8120}
\date{Phys. Rev. B Nov 1994; received 20 June 1994}
\maketitle

\begin{abstract}
Nuclear relaxation in half-integer spin chains at low
temperatures ($T \ll J$,
the antiferromagnetic exchange constant) is dominated by dissipation from
a gas of
thermally-excited, overdamped, spinons.  The universal low temperature
dependence of the
relaxation rates $1/T_1$ and $1/T_{2G}$ is computed.
\end{abstract}
\narrowtext

Nuclear magnetic resonance experiments have recently been shown to
be a powerful tool in studying
the electronic spin dynamics of the two-dimensional quantum
antiferromagnets in the cuprate
compounds~\cite{Imai}.  The combined measurements of the longitudinal
relaxation rate, $1/T_1$,
and the spin-echo decay rate $1/T_{2G}$, over a wide temperature ($T$)
range
allow one to learn a
great deal about the antiferromagnetic spin-fluctuation
spectrum~\cite{Chak-Orbach,Ch-Sach,Sokol,CSS}.

Another class of Heisenberg antiferromagnets with novel properties are
the one-dimensional spin
chains with half-integer spins per site. Some nuclear relaxation
measurements on such systems have been performed~\cite{pincus} and more
detailed studies are under way~\cite{takigawa}.  Our
theoretical understanding of the ground state properties of this system
is in good shape~\cite{affleck}, and there has been limited discussion of
the behavior at finite
temperatures~\cite{schulz,spinless,shankar}.
It is the purpose of this paper to use results on the $T$, wavevector,
and
frequency dependence of the uniform and staggered spin susceptibilities
to obtain the
NMR relaxation rates. There have been earlier
discussions~\cite{pincus,baskaran} of the
$T_1$ relaxation rate in spin-1/2 chains and we will comment on their
relationship to our results below.

An important difference between the two-dimensional antiferromagnets and
the
half-integer spin chains is that the latter are {\em generically\/}
critical.  This means that
the zero-temperature spin correlators have power-law decays in both space
and time, over a finite
range of ratios of short-range exchange interactions - this happens
because the critical
fixed-point has no relevant perturbations which respect the symmetry of
the underlying
Hamiltonian.  However, for sufficiently large second-neighbor coupling
(for example), there is a
transition to a gapped, dimer phase which is not critical.  We will
restrict our attention here
to the range of couplings where the ground state is critical.  This
immediately has important
consequences for the finite temperature spin-correlators:  the entire
low-temperature region ($T
\ll J$) may be considered as the analog of the {\em quantum-critical
region\/} of two-dimensional
antiferromagnets~\cite{CHN,jinwu,Ch-Sach}.  There is now no requirement
that the temperature be
larger than some stiffness-associated energy scale, as there is no analog
of the
renormalized-classical~\cite{CHN} region. Of course, when the small
interchain coupling is taken into account, three-dimensional long-range
order, and the corresponding renormalized-classical region, can appear at
very low temperatures.

The principles of conformal invariance allow one to obtain the exact
scaling functions of the
quantum-critical region of many one-dimensional quantum
systems~\cite{cardy,schulz,spinless,shankar}.
Half-integer spin chains however posses a complicating feature.  There is
a marginally irrelevant
operator at the critical fixed-point, which spoils conformal invariance
by inducing logarithmic
corrections to the leading scaling behavior~\cite{fisher,ziman}.  There
is no analog of this
effect in the two-dimensional antiferromagnets.  To keep the analysis
simple, I will first
discuss the computation in which this marginal operator is ignored.  The
logarithmic corrections
induced by it will be discussed in the next section.

Let us first look at the staggered spin correlations of the half-integer
spin chain.  It is
known~\cite{fisher,ziman} that the equal-time, ground state spin
correlators have the staggered
component
\begin{equation}
\left\langle S_i ( 0 ) S_j ( R) \right\rangle_{T=0} = \delta_{ij} (-1)^R
(\ln
R)^{1/2} \frac{D}{R}
\end{equation}
at large spatial separation $R$.  The constant $D$ is
non-universal and depends upon the choice of microscopic exchange
couplings.  In the absence of
the $\ln R$ term, this result is sufficient to specify the staggered
susceptibility at finite
temperatures~\cite{schulz,spinless,shankar}.
As noted above, we will proceed in the remainder of this section by
ignoring the $\ln R$ term - we will put it back in the next section.
Then, a simple application of the
results of Ref~\cite{shankar} gives us the following result for $\chi_s
(k, \omega)$, the
wavevector ($k$) and frequency dependent staggered susceptibility at
finite $T$ (for $\chi_s$,
$k$ is the deviation of the wavevector from $\pi/a$, where $a$ is the
nearest-neighbor spacing).
\begin{equation} \chi_s ( k, \omega ) = \frac{ D}{2 k_B T} \frac{ \Gamma
\left(
{\displaystyle \frac{1}{4} -i \frac{ \hbar(\omega + c k)}{ 4 \pi k_B T}}
\right) \Gamma \left(
{\displaystyle \frac{1}{4} - i \frac{\hbar(\omega - c k)}{ 4 \pi k_B T}}
\right)}{ \Gamma \left(
{\displaystyle \frac{3}{4} - i \frac{\hbar(\omega + c k)}{ 4 \pi k_B T}}
\right) \Gamma \left(
{\displaystyle \frac{3}{4} - i \frac{\hbar(\omega - c k)}{ 4 \pi k_B T}}
\right)}
\label{chis}
\end{equation}
The quantity $c$ is the $T=0$ spinon velocity. Note that the only
microscopic input
into this result for the susceptibility are the values of $c$ and $D$.
A plot of the spectral function $\mbox{Im} \chi_s (k, \omega) /\omega$
was
presented in Ref~\cite{shankar} for a different value of the critical
exponents;
here we present the imaginary part of the result (\ref{chis})
in Fig~\ref{fig1}.
For small wavevectors, with $\hbar c k$ smaller than or around $k_B T$,
the peak in
the spectral function is at $\omega = 0$, indicating the presence of
overdamped excitations -
spinon excitations interact strongly with other thermally excited
spinons, acquiring a
very short lifetime. We will see below that the NMR relaxation is
dominated by the contribution
of these spinons.
At larger $k$, $\hbar c k \gg k_B T$, the peak in the spectral
function moves to finite $\omega$ (See Fig~\ref{fig1}) - these are
propagating spinons with
a lifetime of order $\hbar / k_B T$.

We turn next to the uniform spin susceptibility, $\chi_u ( k, \omega)$,
where
$k$ is now the true wavevector, measured from the zone center.
Unlike the staggered component, the overall normalization
of the uniform spin susceptibility is not arbitrary, as the total spin is
a conserved
quantity. It is therefore useful to define a magnetization {\em
density\/}
$m_a(R)$ which is the spin per unit length. The $T=0$ correlator of $m$
is
given by~\cite{affleck}
\begin{equation}
\left\langle m_a (R, \tau) m_b ( 0,0 ) \right\rangle_{T=0} = -
\frac{\delta_{ab}}{8 \pi^2}
\left( \frac{1}{(R + i c \tau )^2} + \frac{1}{(R - i c \tau )^2} \right)
\end{equation}
where $\tau$ is the Euclidean time. We can obtain the finite $T$ form of
this correlator
by conformally mapping onto a Matsubara strip, keeping in mind that the
magnetization
density is a component of a current and has non-zero conformal `spin'.
This procedure
yields
\begin{equation}
 \left\langle m_i (R, \tau) m_j ( 0,0 ) \right\rangle = -
\frac{\delta_{ij}}{2}
 \left( \frac{k_B T}{\hbar c} \right)^2 \frac{ \cosh ( 2 \pi R k_B
T/\hbar c)
 \cos ( 2 \pi \tau k_B T/\hbar ) - 1 }{ \left( \cosh ( 2 \pi R k_B T/
\hbar c)
 - \cos ( 2 \pi \tau k_B T/\hbar ) \right)^2}
\label{chiut}
\end{equation}
Performing a Fourier transform of this result to wavevectors $k$ and
Matsubara frequencies
$\omega_n$, we find the remarkably simple result
\begin{equation}
\chi_u ( k , i \omega_n ) = \frac{c}{ 2 \pi \hbar} \frac{k^2 }{c^2 k^2 +
\omega_n^2}
\end{equation}
Note that, unlike $\chi_s (k, \omega )$, all $T$ dependence has
disappeared !
The $T$ dependence in
(\ref{chiut}) is obtained while performing the discrete Fourier transform
from
frequency to time: the $T$ dependence is contained entirely
in the frequency spacing of the
Matsubara sum. Analytically continuing to real frequencies we get
\begin{equation}
\chi_u ( k, \omega ) = \frac{c}{2 \pi \hbar} \frac{k^2 }{c^2 k^2 -
(\omega + i \epsilon)^2}
\label{chiu}
\end{equation}
where $\epsilon$ is a positive infinitesimal. A conspicuous property of
this result is
that there is no damping of the pole at $\omega = ck$, even at finite
$T$. This is, of course,
a property only of the scaling limit. Upon considering corrections to
scaling,
some damping should appear, but will be suppressed by powers of $T/J$.
Related to the absence of damping, is the fact the spectrum of
magnetization fluctuations
is propagating and not diffusive. The spin diffusion constant is
effectively infinite in the
scaling limit.

One might, at this point, raise the issue of whether it is legitimate
to neglect the damping of the uniform magnetization modes: perhaps the
damping coefficient
will be dangerously irrelevant and will contribute a singular $T$
dependence to the
relaxation rates computed below. This possibility appears to me to be
quite unlikely.
The only role of temperature in all of the computations discussed here is
to act as a
finite-size cutoff in the imaginary time direction to a critical theory:
the $T=0$
result should surely be obtained in the limit $T \rightarrow 0$.

We are now finally in a position to compute the nuclear relaxation rates.
We will use the following expressions~\cite{Slichter}, which are
appropriate for the relaxation
of a nucleus coupled to the electronic spins by a hyperfine term:
\begin{eqnarray}
\frac{1}{T_1} &=& \lim_{\omega \rightarrow 0} \frac{ 2 k_B T}{\hbar^2
\omega}
\int \frac{dk}{2 \pi}  A_{\parallel}^2 ( k) \mbox{Im} \chi ( k , \omega )
\nonumber \\
\left(\frac{\hbar}{T_{2G}}\right)^2 &=& \frac{1}{a}~
\int \frac{d k}{ 2 \pi} A^{4}_{\perp} (k) ~
(\chi (k, 0))^2 - \left[\int \frac{d k}{2 \pi}
A^{2}_{\perp} (k) ~ \chi (k, 0)\right]^2
\label{nmr}
\end{eqnarray}
where $a$ is the lattice spacing, and $A_{\parallel}$ ($A_{\perp}$) are
the hyperfine
couplings parallel (perpendicular) to the field; their $k$ dependence is
expected
to be smooth and arises from appropriate form factors.
We have also neglected contributions from nucleus-nucleus dipolar couplings
which could be important in some materials.
The susceptibility
$\chi$ should include contributions from both the uniform and staggered
spin fluctuations.
It is now a straightforward matter to insert (\ref{chis}) and
(\ref{chiu}) into (\ref{nmr})
and obtain the $T$ dependence of the rates. Simple power counting shows
that
the contribution of $\chi_s$ to the rates behaves as $1/T_1 \sim T^{0}$
and $1/T_{2G}
\sim T^{-1/2}$, while the contribution of $\chi_u$ scales as $1/T_1 \sim
T$
and $1/T_{2G} \sim T^{0}$. In both cases the contribution of the
staggered component is
dominant for small $T$. A complete calculation of this term yields
finally the following leading
low $T$ result:
\begin{eqnarray}
\frac{1}{T_1} &=& A_{\parallel}^2 ( \pi/a) \frac{\pi D}{\hbar^2 c}
\nonumber \\
\frac{1}{T_{2G}} &=& A_{\perp}^2 ( \pi/a ) \frac{D}{2\hbar^2 c} \left(
\frac{\hbar c}{k_B T a} \right)^{1/2} I
\label{tres}
\end{eqnarray}
where the numerical factor, $I$ is given by the integral
\begin{equation}
I^2 = \int_0^{\infty} dx \left| \frac{\Gamma ((1 + i x)/4 )}{
\Gamma((3 + i x)/4 )} \right|^4 = 71.276591604 \ldots
\end{equation}
We emphasize that these results have neglected the logarithmic
corrections to scaling
which we will consider below. Note that the unknown prefactor $D$ cancels
out
upon considering the ratio of the rates: this should be a convenient way
of experimentally
testing these results.

The $T$ independent behavior of $1/T_1$ (modulo logarithmic corrections
discussed below) has
already been
noticed in pervious analyses~\cite{pincus,baskaran}; however a different
$T$-independent
value was obtained as these works did not account for the damping of the
spinon states.

\subsection*{Logarithmic corrections}
We now consider the consequence of the marginally irrelevant operator
present in the field
theory of half-integer spin chains~\cite{affleck}. The basic result is
easy to state:
both expressions for the relaxation rates in (\ref{tres}) acquire an
identical, multiplicative
prefactor of $\ln^{1/2} (J/T)$. Further there are subdominant additive
corrections which are
suppressed by powers of $1/\ln (J/T)$: the form of these additive
corrections will be different
for the two relaxation rates. As these additive corrections are only
logarithmically suppressed,
it may be necessary to have $T$ significantly smaller than $J$ before the
leading results
(\ref{tres}) with their $\ln^{1/2} (J/T)$ are accurate.

The arguments for the logarithmic corrections are simple and closely
parallel those presented
in Refs~\cite{fisher,ziman}. One begins by writing down the
Callan-Symanzik equation
for $1/T_1 (T, \Lambda)$, where $\Lambda \sim J$ is an ultraviolet
cutoff. It is known
that this quantity is finite in the limit $\Lambda \rightarrow \infty$
after multiplication
by a $\Lambda$-dependent renormalization factor $Z_{\Lambda}$. This fact
can be used to derive a
Callan-Symanzik equation for $1/T_1$ in which the temperature $T$ scales
under its canonical
dimension as it is nothing but an inverse length in Matsubara time
direction. Integrating
the Callan-Symanzik equation~\cite{fisher,ziman} to a scale where $T \sim
\Lambda$,
expresses $1/T_1$ as $\ln^{1/2} (\Lambda/T)$ times the value of $1/T_1$
in a system
in which the coefficient of the marginally irrelevant coupling $\sim
1/\ln(\Lambda / T)$.
To leading order, we can neglect this coupling and carry out the latter
calculation
in the critical theory with no marginal coupling, which is exactly what
was done above.
A similar argument can be made for $1/T_{2G}$: the $\ln^{1/2} (\Lambda
/T)$ factor
will be the same because the rescaling factor $Z_{\Lambda}$ is identical
to that
for $1/T_1$. However, the perturbative corrections in powers of the
coupling
constant $\sim 1/\ln ( \Lambda / T)$ should be different in the two
rates.

\subsection*{Acknowledgements}
I thank Dr.  Masashi Takigawa of I.B.M.  for useful discussions and
for informing me about his experiments~\cite{takigawa} -
this was directly responsible for the above computation.  I also
benefited from discussions with
G. Baskaran, A. Chubukov and R.  Shankar.
This research was supported by NSF Grant No.  DMR92-24290.

While this manuscript was in preparation, I learned of the work of Eggert
{\em et.
al.\/}~\cite{eggert}:  they computed the low temperature dependence of
the uniform spin
susceptibility $\chi_u $ by methods similar to those used in this paper.

\begin{figure} \caption{A plot of the universal spectral weight
$[(k_B T)^2 / D] \mbox{Im} \chi_s ( k, \omega )/ \hbar \omega$ of the
half-integer spin-chain as
a function of $\overline{\omega} = \hbar \omega / k_B T$ for various
values
of $\overline{k} = \hbar c k / k_B T$. The values for $\overline{k} = 1$
have been scaled down
by a factor of $1/3$. Notice the overdamped peak at $\overline{k}=1$ and
the propagating
peaks at $\overline{k}=3,5$.
}
\label{fig1}
\end{figure}

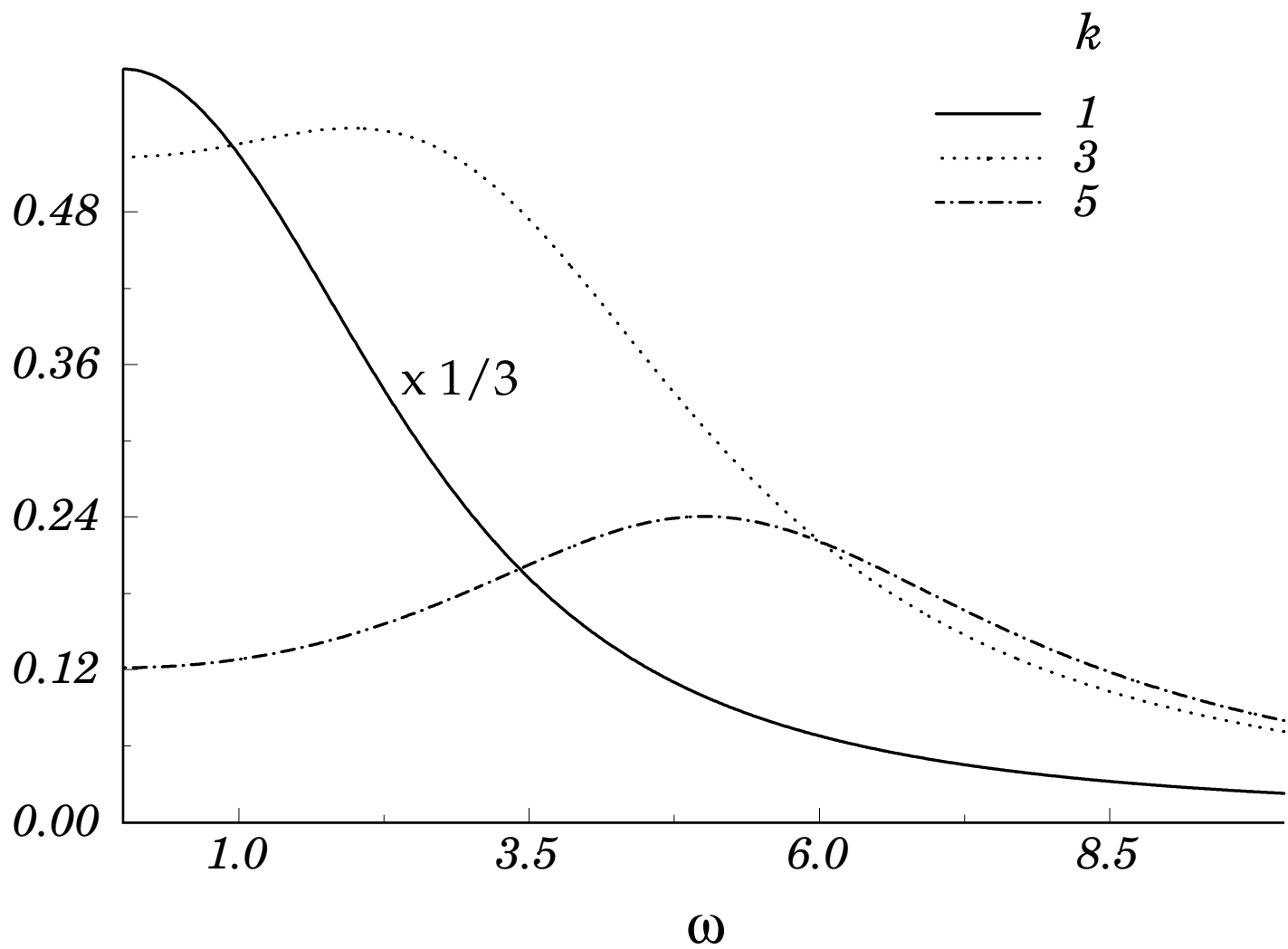
\end{document}